# In-architecture X-ray assisted C-Br dissociation for on-surface fabrication of diamondoid chains

Yan Wang[1,2], Niklas Grabicki[3], Hibiki Orio[4], Juan Li[2], Jie Gao[1,2], Xiaoxi Zhang[1], Tiago F. T. Cerqueira[5], Miguel A. L. Marques[6], Zhaotan Jiang[2], Friedrich Reinert[4], Oliver Dumele[3,7], Carlos-Andres Palma[1]

[1]Institute of Physics, Chinese Academy of Sciences, 100190 Beijing, P. R. China

[2]School of Physics & Advanced Research Institute of Multidisciplinary Science, Beijing Institute of Technology, 100081 Beijing, P. R. China

[3]Department of Chemistry & IRIS Adlershof, Humboldt University of Berlin, 12489 Berlin, Germany

[4]Experimentelle Physik VII and Würzburg-Dresden Cluster of Excellence ct.qmat, Universität Würzburg, 97074 Würzburg, Germany

[5]CFisUC, Department of Physics, University of Coimbra, Rua Larga, 3004-516 Coimbra, Portugal

[6]Research Center Future Energy Materials and Systems of the University Alliance Ruhr and Interdisciplinary Centre for Advanced Materials Simulation, Ruhr University Bochum, Universitätsstraße 150, D-44801 Bochum, Germany

[7]Institute of Organic Chemistry, University of Freiburg, Albertstrasse 21, 79104 Freiburg, Germany

**ABSTRACT**

The fabrication of well-defined, low-dimensional diamondoid-based materials is a promising approach for tailoring diamond properties such as superconductivity. On-surface self-assembly of halogenated diamondoids under ultrahigh vacuum conditions represents an effective strategy in this direction, enabling reactivity exploration and on-surface synthesis approaches. Here we demonstrate through scanning probe microscopy, time-of-flight mass spectrometry and photoelectron spectroscopy, that self-assembled layers of dibromodiamantanes on gold can be debrominated at atomic wavelengths (Al Kα at 8.87 Å and Mg Kα at 9.89 Å) and low temperatures without affecting their well-defined arrangement. The resulting 'in-architecture' debromination enables the fabrication of diamantane chains from self-assembled precursors in close proximity, which is otherwise inaccessible through annealing on metal surfaces. Our work introduces a novel approach for the fabrication of nanodiamond chains, with significant implications for in-architecture and layer-by-layer synthesis.

**INTRODUCTION**

Well-defined diamondoid chains are extended $sp^3$-carbon nanomaterials expected to feature the excellent properties of nanodiamond units, such as mechanical hardness, stability, thermal conductivity, strong electron-phonon coupling and corresponding optical properties[1-11]. In the past years, several families of diamondoid-containing nanomaterials have been introduced. These include diamondoids bridged by metals or heteroatoms on Cu(111) and Au(111) surfaces[12, 13], C-C coupled diamantanes within carbon nanotubes[14], diamondoid monolayers[15] and field emitters[16], fullerene-diamantane hybrid rectifier[17], and diamond nanothreads with excellent mechanical properties. To date, the growth and characterization of a single, atomically-precise diamondoid chain remains to be explored.

A promising approach granting access to concomitant fabrication and physical property characterization of diamondoid chains is the adsorption of appropriate precursors on metal surfaces and their subsequent on-surface polymerization under ultra-high vacuum (UHV) conditions. Scanning probe microscopy studies of halogen-based precursors are a well-established route to this end[18-20], aiming at C-C Ullmann-like coupling on metal surfaces. There are essentially two methods for activating halogen-based C-C coupling: Catalytic thermal dehalogenation and stimuli-induced dehalogenation which includes photo-[21-26] or electron-source induced[22, 25] dehalogenation. Following dehalogenation, molecular radicals incur diffusion and stabilization under UHV conditions, especially on metal surfaces[14, 27]. This grants access to efficient radical-radical coupling reactions to form C-C bonds[28, 29] on metal surfaces, provided adequate reaction rates and desorption suppression at the dehalogenation temperatures, a condition usually met by flat, strongly-adsorbed $sp^2$-carbon polycyclic aromatic hydrocarbons[30]. However, for three-dimensional, weakly-adsorbed molecules such as diamantane, desorption is expected to occur during thermal dehalogenation conditions under UHV conditions, prompting the development of alternative strategies for dehalogenation and coupling. One strategy for preventing desorption and promoting proximal C-C coupling is topochemistry, which involves inducing chemical reactions between neighboring molecules within an assembly[31]. Diverse topochemical conditions have been suggested to occur via proximity-enabled radical-radical coupling in the dehalogenation of molecular crystals[23, 25]. Such a strategy would ideally rely on external stimuli such as electrons from an STM tip or photon sources[32]. Unlike benchtop photon sources, synchrotron radiation would expand the accessible mechanism of photo-induced dissociation, from valence-band excitation to dissociation channels to core-level excitations. Additionally, employing atomic-scale wavelengths and in-architecture coupling would enable the photopatterning of complex computing and functional architectures at the ultimate miniaturization limit[25, 33]. More importantly, localized radical species could be brought into contact with other reactive atomic or molecular species, defining a general method for proximity-enabled, layer-by-layer fabrication. Notwithstanding several works indicating dehalogenation in self-assembled monolayers and crystals to achieve hierarchical reactivity[34], direct evidence of in-architecture radical formation, for proximity-enabled radical-radical coupling within a self-assembled architecture, that is, in-architecture coupling, is limited.

Herein, we demonstrate the X-ray induced debromination of 4,9-dibromodiamantane self-assembled architectures on Au(111) surface (**Figure 1**) employing an Al Kα at 8.87 Å ($h\nu$ = 1486.6 eV) source. Low-temperature scanning tunneling microscopy (LT-STM at 77 K and 4 K) is employed to characterize the surface configuration of molecules and their transformation before and after X-ray irradiation at low temperatures. X-ray photoelectron spectroscopy (XPS) and time-of-flight mass spectrometry (TOF-MS) is performed to compare the differences between annealing and X-ray induced C-Br cleavage. These joint measurements reveal the self-assembly of $sp^3$-carbon nanostructures on metal surfaces, introduce the chemical modification effects of X-ray irradiation on $sp^3$-carbon nanostructures, and point out the potential for constructing diamantane dimers through radical coupling using atomic-scale wavelengths. Our work is relevant for advanced photolithography for diamantane in-architecture fabrication with atomic-resolution, as well as layer-by-layer in-architecture synthesis, doping or atom-exchange.

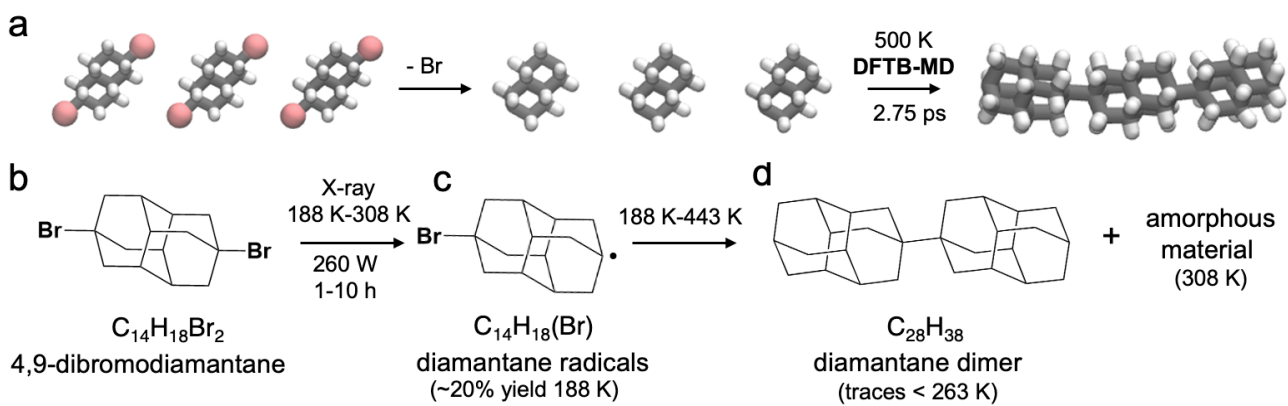

**Figure 1. Schematic diagram of molecular reactions. a**. DFTB-MD simulation of reaction for the synthesis of linear diamantane chain through radical coupling. The optimized (debrominated) radical diamantane molecule system is kept under the temperature of 500K, and after a period of time of around 2.75 ps, a linear chain of diamantane is observed. **b**. Structure of 4,9-dibromodiamantane precursor with size of 8.5 Å. **c**. Structure of diamantane biradical after Al Kα at 8.87 Å source X-ray irradiation. **d**. Structure of diamantane dimer via diamantane radical coupling with size of 13.3 Å, analyzed by STM.

## RESULTS

### Modeling of diamantane coupling

Diamondoids on metal surfaces are characterized by their bulkiness, rigidity, and incompressible 3D structure[35, 36], and the self-assembly of these bulky compounds is mostly stabilized by London dispersion interactions[36]. Closely-packed radical diamondoids (**Figure 1a**) appear to undergo radical-radical coupling in molecular dynamics simulations by means of density functional tight binding (DFTB, see Methods) yielding polydiamantane. Undoped structures of polydiamantane appear to be dynamically stable, while weakly-doped hole-doped structures develop smooth imaginary frequencies at low phonon wavevectors. Thus, charge density waves are expected to evolve in weakly-interacting crystals and lattices of covalently-coupled diamantanes.

### Self-assembly of 4,9-dibromodiamantane and bromodiamantane

The self-assembly of 4,9-dibromodiamantane molecules (see Methods, **Figure S1-S4**) on Au(111) is first investigated via STM under UHV conditions (**Figure 2**) revealing two different domains in the self-assembly of dibromodiamantane on Au(111) surface. The homogeneous self-assembly of molecules in phase 1 with a constant of i **a** = 7.9 ± 0.1 Å, **b** = 8.3 ± 0.1 Å, the lattice constant of phase is illustrated in **Figure 2a, c**. Individual molecules exhibit a pear-like shape, with the circular features visible in the upper left region of the molecules in **Figure 2c** attributed to bromine atoms, while the elliptical form corresponds to the diamantane cage structure. This interpretation considers that the second bromine atom points towards the gold substrate. To corroborate this, monobromodiamantane monolayers are equally studied, and a shorter neighbouring distance is found of 7.4 ± 0.1 Å corresponding to two standing-up monobromodiamantanes (7.1 Å simulated, **Figure S5**).

In the depicted phase 1, dibromodiamantane molecules are tilted in the same direction, as shown in the black rectangle. The proposed configurations of dibromodiamantane on Au(111) are presented in **Figure 2e**, where the top bromine atoms are identified by red spheres. The experimentally measured distance between two molecules is close to 8.3 Å (**Figure 2c**), while the theoretical distance in the proposed atomistic model is 8.7 Å. **Figure S6a** and **S6b** depicts further data on an additionally observed phase, designated as phase 2, with a lattice constant $c$ = 8.1 ± 0.1 Å, $d$ = 12.8 ± 0.1 Å. The vacancy observed in **Figure S6b**, referred to as a "1,2-vacancy," arises from a change in the alignment direction of the molecules and is magnified in **Figure S6c**. The proposed configuration to explain the "1,2-vacancy" on the surface is shown in **Figure S6d**.

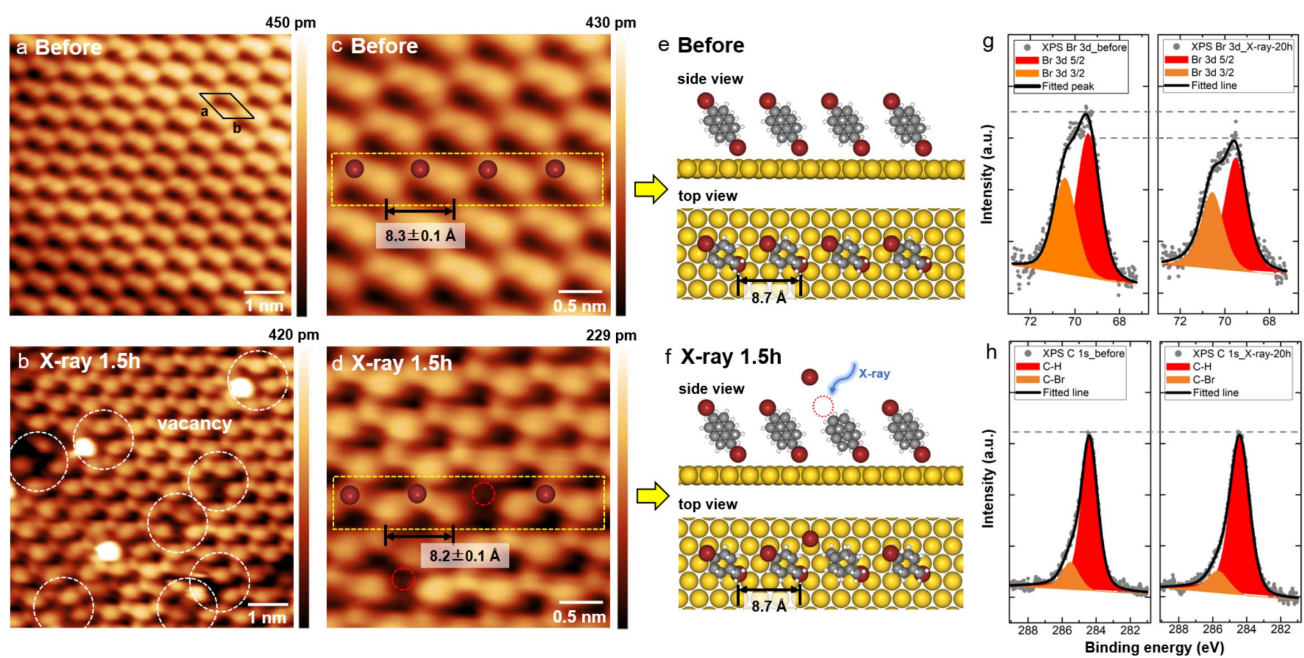

**Figure 2. Dibromodiamantane on Au(111) before and after X-ray irradiation 218-263 K. a.** STM image of the self-assembly of dibromodiamantane molecules on Au(111), the unit cell is **a** = 7.9 Å ± 0.1 Å, **b** = 8.3 Å ± 0.1 Å. (0.8 V, 35 pA). **b**. STM data of dibromodiamantane molecules on Au(111) after X-ray irradiation for 1.5 hours (0.8 V, 20 pA). **c.** Magnified STM image of the self-assembly on Au(111) (0.8 V, 35 pA). **d.** Magnified STM data of dibromodiamantane molecules on Au(111) after X-ray irradiation Al Kα at 8.87 Å for 1.5 hours (0.8 V, 20 pA). **e.** Proposed spatial configurations of dibromodiamantane molecules on Au(111), corresponding to the molecules marked by the yellow rectangle in panel c. **f.** Proposed spatial configurations of dibromodiamantane molecules on Au(111) after X-ray irradiation at 223 K to 263 K, corresponding to the molecules marked by the yellow rectangle in panel d. **g**. XPS spectra on Br 3d of dibromodiamantane before and after X-ray irradiation for 20 hours with Mg Kα 9.89 Å ($h\nu$ = 1253.6 eV). **h**. XPS spectra on C 1s of dibromodiamantane before and after X-ray irradiation for 20 hours at 218 K with Mg Kα 9.89 Å ($h\nu$ = 1253.6 eV).

## Low-efficiency X-ray assisted bromine dissociation between 218-263 K

Dibromodiamantane samples are subsequently exposed to X-ray irradiation for 1.5 hours at temperature from 223 to 263 K. Defects, such as vacancies and bright protrusions, were observed to cover less than 20% of the irradiated samples. The limited extension of defects correlates with the low reactivity of the monolayer towards Al Kα at 8.87 Å source, as revealed by XPS studies conducted at 218 K. The domains display Y-shaped defects, marked with dashed circles in **Figure 2b**. The vacancies are assumed to be caused by the dissociation of bromine atoms from the molecules. A magnified STM image of a vacancy, marked by a red dotted line, is shown in **Figure 2d**. The proposed spatial configurations explaining the surface structure, highlighted by a yellow rectangle in **Figure 2d**, are provided in **Figure 2f**, where a vacancy is attributed to the loss of a bromine atom. The experimentally measured distance between adjacent molecules is approximately 8.2 Å (**Figure 2d**), close to the theoretical distance of 8.7 Å in the proposed molecular model.

XPS is performed on the samples before and after 20 hours of X-ray irradiation, with the irradiated samples maintained at a lower temperature of 218 K. The analysis first focus on the Br 3d region, as shown in **Figure 2g**. The fitted peaks for Br 3d peak reveal binding energies of Br $3d_{5/2}$ = 69.50 eV and Br $3d_{3/2}$ = 70.57 eV, which are typical values for bromine in a covalent bond with carbon[22, 24, 37]. After X-ray irradiation, the intensity of the Br 3d peak decrease by 17%, further corroborating the occurrence of debromination. Additionally, the C 1s region is investigated to determine whether the reduction in bromine atoms following X-ray irradiation is due to the cleavage of the C-Br bond or the desorption of molecules from the substrate. As illustrated in **Figure 2h**, the intensity of the C1s peak is negligibly changed before and after X-ray irradiation. The peak at 285.52 eV, corresponding to the C-Br bond[38], show a 14% decrease in area after irradiation, while the peak at 284.39 eV, associated with the C-H bond[12], exhibit a 15% increase in area. These observations indicate that the change in the Br 3d peak intensity is attributable to the breaking of the C-Br bond.

Populating bromine antibonding orbitals was previously found to contribute to dehalogenative dissociation channels[22]. Similar antibonding excitation pathways could be promoted through X-rays through Br M, L-edge absorption, Auger and secondary electrons, all of which are possible at the employed photon energy. The poor dehalogenation yield after 20 hours of irradiation at low temperatures tentatively points towards very short-lived excited states which might require either higher photon fluxes or increasing the absorption cross-section to enhance the probability of phonon-activation.

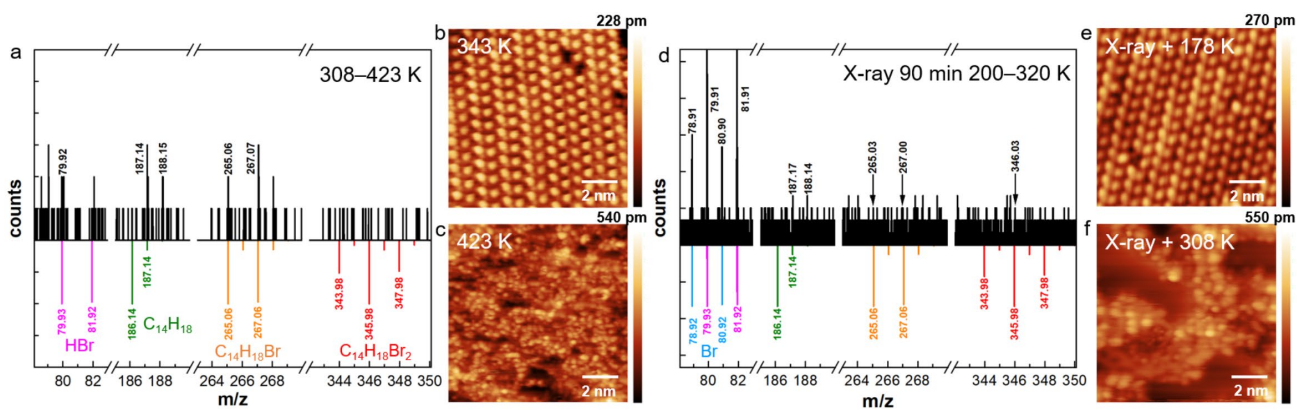

**Figure 3. Annealing to 343 and 423 K vs. X-ray irradiation of dibromodiamantane at 308 K. a.** 30 min-TOF-MS of dibromodiamantane while TPD annealing from 308 K to 423 K, and HBr, diamantane biradical, diamantane-Br radical, dibromodiamantane simulation peaks. **b-c.** STM images of dibromodiamantane after annealing to 343 K and 423 K, respectively. Parameters: $V_b$ = 0.8 V, $I_t$ = 25 pA, $T_{STM}$ = 77 K. **d.** 90min-TOF-MS of dibromodiamantane while X-ray irradiation from 200 K to 320 K, and HBr, diamantane biradical, diamantane-Br radical, dibromodiamantane simulation peaks. **e-f.** STM images of dibromodiamantane after X-ray irradiation at sample temperatures of 178 K and 308 K, respectively. Parameters: e. $V_b$ = 0.8 V, $I_t$ = 35 pA, $T_{STM}$ = 77 K. f. $V_b$ = 0.8 V, $I_t$ = 25 pA, $T_{STM}$ = 77 K.

**X-ray damage at ~308 K vs. thermal dissociation**

Thermal debromination reaction on Au(111) is reported to occur between 473 K and 523 K[23, 24]. When annealing to promote thermal debromination, we observe increasingly disordered material at 343 K. Complete desorption of the material is found after flash annealing to 423 K (**Figure 3a**), at which point an adlayer of disordered contaminants is observed and assigned to bromine. The situation drastically changes after X-ray irradiation for 7 hours at ~308 K where patterns reminiscent of damaged material are found (**Figure 3b**).[14] To assess whether areas of damaged material are prone to chain-building, samples are further studied by room temperature STM. When scanning at room temperature, disordered material is plausibly highly mobile and not imaged, and rather elongated adsorbates are observed (**Figure S10**).

To compare thermal and X-ray induced debromination in detail, TOF-MS is acquired by means of 10 eV electron-impact ionization (EI) on samples annealed to 423 K without and to 308 K with X-ray irradiation. As shown in **Figure 3c**, TOF-MS of dibromodiamantane samples without X-ray irradiation during desorption ~308-423 K, identify several peaks of molecular fragments. The peak at $m/z$ = 79.92, is assigned to HBr$^+$ (theoretical *m/z* = 79.93), indicating bromine dissociation as a result of either thermal annealing or EI ionization. Peaks at *m/z* = 187.14, 188.15 are close to C$_{14}$H$_{19}$$^{•+}$ (theoretical *m/z* = 187.15, 188.15), showing dibromodiamantane molecules (C$_{14}$H$_{18}$Br$_2$) lose two bromine atoms during measurement and gain a hydrogen, similar to HBr$^+$. Peaks at *m/z* = 265.06, 267.07 are in agreement with C$_{14}$H$_{18}$Br$^{•+}$ (theoretical *m/z* = 265.06, 267.07), demonstrating dibromodiamantane molecules (C$_{14}$H$_{18}$Br$_2$) may lose one bromine during TOF-MS measurement. It's worth noting that there is no dibromodiamantane peak (theoretical *m/z* = 345.98) in the data, indicating that dibromodiamantane is either thermally unstable while annealing to 423 K, or broken by 10 eV EI during TOF-MS measurement. For the TOF-MS while warming up the sample from 223 K to ~330 K under 90 minutes of X-ray irradiation (**Figure 3d**), fewer signals are observed. One notable difference are signals assigned to Br$^+$ at m/z = 80.92 and 78.92. One interpretation for the measurement of Br$^+$ is the desorption of Br due to X-ray assisted dissociation of Br from the dibromodiamantane monolayer. Another interpretation is X-ray assisted desorption of Br from Au(111). We note that the acquisition of TOF-MS was performed every 15 minutes during the 90 min irradiation (the plots are the sum of six spectra).

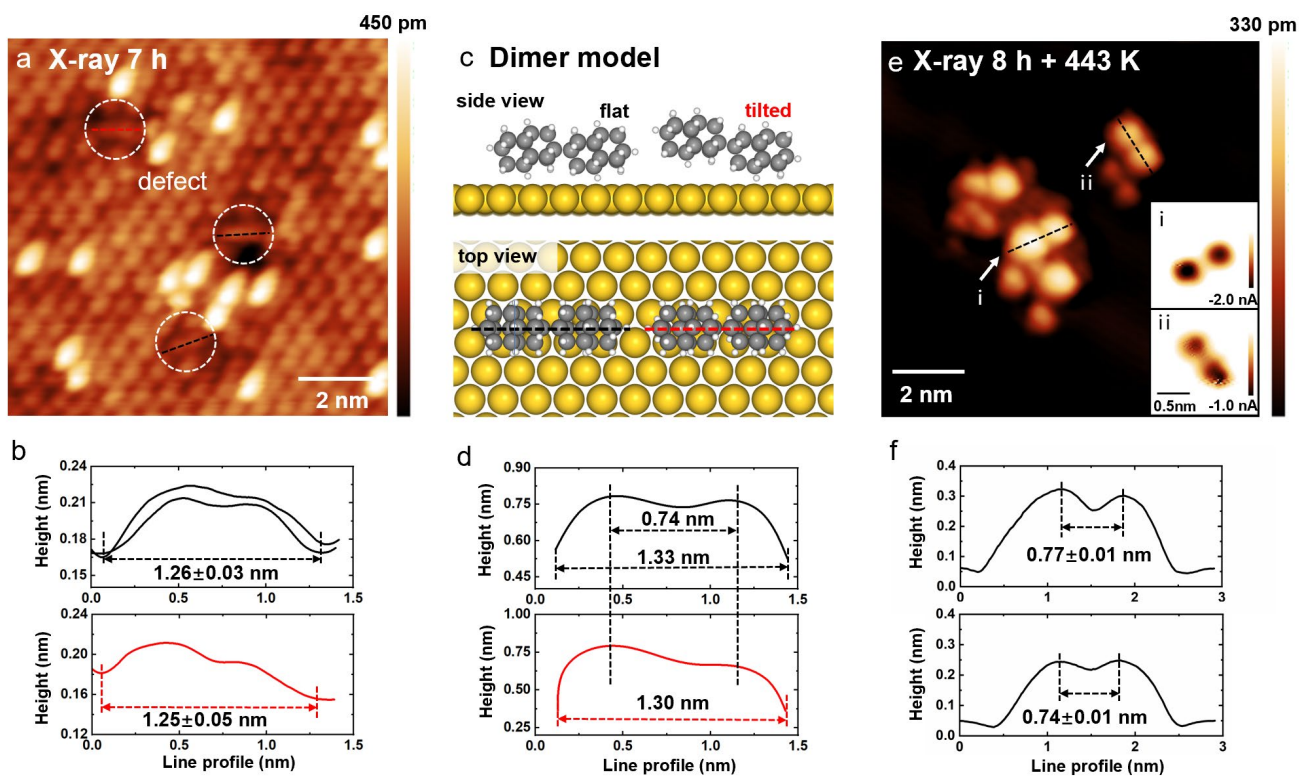

**Figure 4. Dibromodiamantane monolayer on Au(111) after 7 hours of X-ray irradiation at ~250 K. a**. STM data of dibromodiamantane molecules on Au(111) after X-ray irradiation for 7 hours (0.8 V, 20 pA) from 220 K to 267, $T_{STM}$ = 77 K. **b**. Line profiles of the defects over the sample surface in panel a, the black are corresponding to the black lines in panel a, the red is corresponding to the red line in panel a. **c**. Proposed diamantane dimer model on Au(111). **d**. Simulation size of the diamantane dimer in panel b, the black corresponds to the flat dimer in panel b, the red corresponds to the tilted dimer in panel b. **e**. STM data of X-ray irradiation for 8 hours and 443 K annealing, $T_{STM}$ = 4 K. The inset is the constant-height data of the two dimer-like features i and ii. **f**. Line profiles of dimers over sample surface in panel e.

**Optimization of X-ray assisted debromination at ~220 to 263 K for in-architecture synthesis**

When increasing the duration of X-ray irradiation to ~7 hours, STM data reveals a pronounced defect density as the temperature approaches ~250 K. **Figure 4a** shows the result of irradiation and temperature optimization, whereby additional bright protrusions are found when compared to **Figure 2**. In addition, elongated defects appear on the surface, marked by white dotted lines. The line profile over these defects in **Figure 4c** reveals two similar profiles, with a length of approximately ~1.25 nm. To understand the structure of these defects, the diamantane dimer with the corresponding spatial configurations on Au(111) is modeled in **Figure 4b**. The models indicate total hydrogen-to-hydrogen lengths ranging from 1.33 nm to 1.30 nm (**Figure 4d**), which might hint at the formation of diamantane dimers. Further annealing of the sample to desorb unreacted material to 443 K leads to the observation of protrusions of comparable length to the diamantane dimer models (**Figure 4e** and **4f**), albeit surrounded by unidentified material. We attribute the presence of undefined species to the annealing process concentrating both, the putative reacted diamantane material with higher sublimation temperature, and contaminants–noting that the samples were exposed to X-ray irradiation for 7 hours at ~250 K and pressures of ~$10^{-9}$ mbar.

The phonon spectrum of polydiamantane has been further calculated, as shown in **Figure S11**. The results indicate that while the undoped structures are dynamically stable (**Figure S11b**), the hole-doped structures exhibit smooth imaginary frequencies at low phonon wavevectors (**Figure S11d**). This suggests the potential emergence of charge density waves in weakly interacting crystals and lattices of polydiamantane.

**SUMMARY**

Benchtop X-ray irradiation was found to induce distinct types of defects in diamantane monolayers at temperatures up to 263 K, one of which was assigned to C-Br dissociation. At temperatures close to 300 K, X-ray irradiation incurs damage and desorption of monolayers. In contrast, monolayers that were not subjected to irradiation remain intact at least up to 343 K. Long irradiation times at non-damaging temperatures of approximately 250 K enhance C-Br dissociation whereby potential dimers of diamantane were identified. Tentative dimers may form through an in-architecture fabrication mechanism, by which neighboring diamantane radicals react. Our work serves as a stepping stone towards synchrotron studies for on-surface synthesis aiming at resonant X-ray dissociation, while probing the ultimate single-atom dissociation photolithography limit.

**METHODS**

**Sample preparation**. Au(111)/mica substrate was sputtered by argon ions using 1 kV and 10 mA under $1.0 \times 10^{-5}$ mbar, then annealing to 723 K. Dibromodiamantane molecules were deposited on Au(111) from a glass crucible held at room temperature keeping 98 minutes under pressure from $3.4 \times 10^{-9}$ mbar to $2.8 \times 10^{-9}$ mbar, while the Au(111) was kept at 178 K during deposition. After deposition, samples are irradiated by X-ray source, the distance between the sample and X-ray was 1.5 cm. First, the dibromodiamantane samples were irradiated by X-ray with 260 W for 1.5 hours, the sample temperature was warmed naturally from 223 K to 263 K during irradiation, the pressure was changing from $5.7 \times 10^{-9}$ mbar to $1.0 \times 10^{-8}$ mbar during irradiation. Second, the dibromodiamantane samples were further irradiated by X-ray with 260 W for 2.5 hours (sum to 4 hours), the sample temperature was warmed naturally from 188 K to 263 K during irradiation, the pressure was changed from $5.7 \times 10^{-9}$ mbar to $7.5 \times 10^{-9}$ mbar during irradiation. Third, the dibromodiamantane samples were further irradiated by X-ray with 260 W for 3 hours (sum to 7 hours), the sample temperature was warmed naturally from 208 K to 263 K during irradiation, the pressure was changed from $5.6 \times 10^{-9}$ mbar to $6.5 \times 10^{-9}$ mbar during irradiation. Fourth, the dibromodiamantane samples were further irradiated by X-ray with 260 W for 3 hours (sum to 10 hours), the sample temperature was warmed naturally from 198 K to 278 K during irradiation, the pressure was changed from $5.4 \times 10^{-9}$ mbar to $7.3 \times 10^{-9}$ mbar during irradiation. Monobromodiamantane molecules were deposited on Au(111) from a glass crucible held at room temperature keeping 5 minutes under $1.0 \times 10^{-8}$ mbar, and the Au(111) was kept at 178 K during deposition. After deposition, samples were irradiated by X-ray Al Kα at 8.87 Å, the distance between the sample and X-ray was 1.5 cm at a 45 °C angle. The monobromodiamantane samples were irradiated by X-ray with 260 W for 1.5 hours, the sample temperature was kept at 178 K during irradiation, the pressure was changed from $3.4 \times 10^{-9}$ mbar to $4.4 \times 10^{-9}$ mbar during irradiation.

**Scanning tunneling microscopy**. Scanning tunneling microscopy (STM) was carried out using a *CreaTec GmbH* 1 K SPM operating at 77 K or 4 K under ultra-high vacuum conditions. The Pt/Ir tip was used for STM, and all STM data were obtained in either constant current or feedback-off mode, the STM images were processed by Gwyddion[39] software.

**Time-of-flight mass spectrometry**. Time-of-flight mass spectrometry (TOF-MS) was performed using a *Kore-TOF-MS* under ultra-high vacuum conditions, it has an electron impact ion source working at 10 eV ionization energy and 200 μA emission current. All TOF-MS data were processed by *mMass*[40] software, and calibrated by the fitted line obtained from the signals of $H_2O$, $Si_3O_3C_5H_{15}$, $Si_4O_4C_7H_{21}$ present in every sample.

**X-ray photoelectron spectroscopy.** Au(111) single crystal was cleaned by Ar sputtering and then annealed at 873 K. The cycle was repeated several times. Molecules were deposited on an Au(111) single-crystal substrate from a Knudsen cell held at room temperature and the substrate was kept at 138-177 K during the deposition. XPS was performed in an ultra-high vacuum chamber (base pressure below $4 \times 10^{-10}$ mbar) using a hemispherical electron-energy analyzer (R3000, SIENTA Omicron) at a sample temperature of 218 K. The emission angle of photoelectron was 60° against sample surface. The X-ray source was unmonochromatized Mg K$\alpha_1$ 9.89 Å ($h\nu$ = 1253.6 eV) at a power of 225 W and had a spot-size of approximately 5 mm in diameter. The sample current was 82 nA. We estimated a photon density of $2.0 \times 10^{12}$ photons·cm$^{-2}$·s$^{-1}$ using assumption that one photon produces one electron. The XPS data were normalized by data acquisition time. We used a combination of pseudo-Voigt functions and the Shirley background as the fitting model. The two peak widths of Br $3d_{5/2}$ and Br $3d_{5/2}$ were constrained to be equal. The same constraint applied to the widths of C-H and C-Br. For the Br 3d spectra, the contributions from Au $4f_{5/2}$ satellite peak at 70.6 eV due to Mg K$\alpha_5$ ($h\nu$ = 1271.0 eV) was subtracted.

**DFT simulation.** All density-functional calculations are performed using Quantum Espresso[41, 42], with the Perdew-Burke-Ernzerhof (PBE) for solids (PBEsol)[43] generalized gradient approximation. Geometry optimizations are performed using a uniform gamma-centered 4x4x4 k-point grid. Convergence thresholds for energies, forces and stresses are set to 1E-8 a.u., 1E-6 a.u., and 5E-2 kbar, respectively. Phonon frequencies are computed on a 8x8x8 k-grid and a 2x2x2 q-grid within density functional perturbation theory[44]. We use pseudopotentials from the pseudodojo project[45], specifically the stringent norm-conserving set. The plane-wave expansion cutoff is set at 90 Ry.

**Molecular Dynamics simulation.** The DFTB+[46] software package had been used to carry out the optimization and molecular dynamical study of the diamantine chains. The driver used in the optimization process was Conjugate Gradient with Hamiltonian as DFTB (DFTB uses approximate waves functions to calculate the most stable electronic structure). The Slater Koster files used were 3ob-3-1 with suitable periodic boundary conditions (PBC) & K-points considering supercell samplings. The DFTB Hamiltonian was used in all of the MD processes making it a DFTB-MD with Velocity

Verlet algorithm used as a driver. The Nosehoover thermostat algorithm had been used to make a constant temperature system i.e NVT simulation with using the same Slater-Koster file as used in optimization i.e 3ob-3-1 with PBC and K-points supercell samplings. The coupling strength had been maintained at 3000 cm$^{-1}$ in all cases. The optimized debrominated diamantane molecule system was kept at the temperature of 500 K. As the simulation began, we observed the linkage of diamantanes as dimers & trimers. After a period of time around (2.75 ps), a linear chain of around ten diamantanes was observed. From this simulation, we could ensure the linear-chain diamantane formation experimentally.

## ACKNOWLEDGMENTS

This work was financially funded by the National Natural Science Foundation of China (nos. 11974403, 12474178, 12304238), Beijing Natural Science Foundation (nos. IS24031), Sino-German Project (nos. 51761135130), the Chinese Academy of Sciences (nos. QYZDBSSW-SLH038, XDB33000000), China Postdoctoral Science Foundation (2023M733721, 2024T170991), the Max Planck Society, the European Commission through the Graphene Flagship and the FET Open 2D Ink Project (nos. 664878), and the Würzburg-Dresden Cluster of Excellence on Complexity and Topology in Quantum Matter - ct.qmat (EXC 2147, project-id 390858490). We gratefully acknowledge financial support from the Alexander von Humboldt Foundation (C. -A. P.) and hosting by Norbert Koch and Jürgen P. Rabe.

## ASSOCIATED CONTENT

**Supporting Information**

The Supporting Information is available free of charge.

## AUTHOR INFORMATION

**Corresponding authors**

**Notes**

The authors declare no competing interests.

# References

1. Marchand, A. P., Diamondoid Hydrocarbons--Delving into Nature's Bounty. *Science* **2003**, *299* (5603), 52-53.
2. Schwertfeger, H.; Fokin, A. A.; Schreiner, P. R., Diamonds are a chemist's best friend: diamondoid chemistry beyond adamantane. *Angew. Chem. Int. Ed.* **2008**, *47* (6), 1022-1036.
3. Clay, W. A.; Dahl, J.; Carlson, R.; Melosh, N.; Shen, Z., Physical properties of materials derived from diamondoid molecules. *Rep. Prog. Phys.* **2014**, *78* (1), 016501.
4. Dahl, J.; Liu, S.; Carlson, R., Isolation and structure of higher diamondoids, nanometer-sized diamond molecules. *Science* **2003**, *299* (5603), 96-99.
5. Mansoori, G. A., Diamondoid molecules. *Adv. Chem. Phys.* **2008**, *136*, 207-58.
6. Yan, H.; Hohman, J. N.; Li, F. H.; Jia, C.; Solis-Ibarra, D.; Wu, B.; Dahl, J. E.; Carlson, R. M.; Tkachenko, B. A.; Fokin, A. A., Hybrid metal–organic chalcogenide nanowires with electrically conductive inorganic core through diamondoid-directed assembly. *Nat. Mater.* **2017**, *16* (3), 349-355.
7. Mochalin, V. N.; Shenderova, O.; Ho, D.; Gogotsi, Y., The properties and applications of nanodiamonds. *Nat. Nanotechnol.* **2012**, *7* (1), 11-23.
8. Patrick, C. E.; Giustino, F., Quantum nuclear dynamics in the photophysics of diamondoids. *Nat. Commun.* **2013**, *4* (1), 2006.
9. Filik, J.; Harvey, J. N.; Allan, N. L.; May, P. W.; Dahl, J. E.; Liu, S.; Carlson, R. M., Raman spectroscopy of diamondoids. *Spectrochim. Acta A* **2006**, *64* (3), 681-692.
10. Landt, L.; Klünder, K.; Dahl, J. E.; Carlson, R. M.; Möller, T.; Bostedt, C., Optical response of diamond nanocrystals as a function of particle size, shape, and symmetry. *Phys. Rev. Lett.* **2009**, *103* (4), 047402.
11. Zhou, Y.; Brittain, A. D.; Kong, D.; Xiao, M.; Meng, Y.; Sun, L., Derivatization of diamondoids for functional applications. *J. Mater. Chem. C* **2015**, *3* (27), 6947-6961.
12. Gao, H.-Y.; Sekutor, M.; Liu, L.; Timmer, A.; Schreyer, H.; Mönig, H.; Amirjalayer, S.; Fokina, N. A.; Studer, A.; Schreiner, P. R., Diamantane suspended single copper atoms. *J. Am. Chem. Soc.* **2018**, *141* (1), 315-322.
13. Feng, K.; Solel, E.; Schreiner, P. R.; Fuchs, H.; Gao, H.-Y., Diamantanethiols on metal surfaces: spatial configurations, bond dissociations, and polymerization. *J. Phys. Chem. Lett.* **2021**, *12* (13), 3468-3475.
14. Nakanishi, Y.; Omachi, H.; Fokina, N. A.; Schreiner, P. R.; Kitaura, R.; Dahl, J. E.; Carlson, R. M.; Shinohara, H., Template synthesis of linear‐chain nanodiamonds inside carbon nanotubes from bridgehead‐halogenated diamantane precursors. *Angew. Chem. Int. Ed.* **2015**, *54* (37), 10802-10806.
15. Yang, W. L.; Fabbri, J.; Willey, T.; Lee, J.; Dahl, J.; Carlson, R.; Schreiner, P.; Fokin, A.; Tkachenko, B.; Fokina, N., Monochromatic electron photoemission from diamondoid monolayers. *Science* **2007**, *316* (5830), 1460-1462.
16. Narasimha, K. T.; Ge, C.; Fabbri, J. D.; Clay, W.; Tkachenko, B. A.; Fokin, A. A.; Schreiner, P. R.; Dahl, J. E.; Carlson, R. M.; Shen, Z., Ultralow effective work function surfaces using diamondoid monolayers. *Nat. Nanotechnol.* **2016**, *11* (3), 267-272.
17. Randel, J. C.; Niestemski, F. C.; Botello-Mendez, A. R.; Mar, W.; Ndabashimiye, G.; Melinte, S.; Dahl, J. E.; Carlson, R. M.; Butova, E. D.; Fokin, A. A., Unconventional molecule-resolved current rectification in diamondoid–fullerene hybrids. *Nat. Commun.* **2014**, *5* (1), 4877.
18. Bieri, M.; Treier, M.; Cai, J.; Aït-Mansour, K.; Ruffieux, P.; Gröning, O.; Gröning, P.; Kastler, M.; Rieger, R.; Feng, X.; Müllen, K.; Fasel, R., Porous graphenes: two-dimensional polymer synthesis with atomic precision. *Chem. Commun.* **2009**, (45), 6919-6921.
19. Grill, L.; Hecht, S., Covalent on-surface polymerization. *Nat. Chem.* **2020**, *12* (2), 115-130.
20. Cai, J.; Ruffieux, P.; Jaafar, R.; Bieri, M.; Braun, T.; Blankenburg, S.; Muoth, M.; Seitsonen, A. P.; Saleh, M.; Feng, X.; Müllen, K.; Fasel, R., Atomically precise bottom-up fabrication of graphene nanoribbons.

*Nature* **2010,** *466* (7305), 470-473.

21. Nacci, C.; Schied, M.; Civita, D.; Magnano, E.; Nappini, S.; Píš, I.; Grill, L., Thermal- vs Light-Induced On-Surface Polymerization. *J. Phys. Chem. C* **2021,** *125* (41), 22554-22561.
22. Palma, C.-A.; Diller, K.; Berger, R.; Welle, A.; Björk, J.; Cabellos, J. L.; Mowbray, D. J.; Papageorgiou, A. C.; Ivleva, N. P.; Matich, S.; Margapoti, E.; Niessner, R.; Menges, B.; Reichert, J.; Feng, X.; Räder, H. J.; Klappenberger, F.; Rubio, A.; Müllen, K.; Barth, J. V., Photoinduced C–C Reactions on Insulators toward Photolithography of Graphene Nanoarchitectures. *J. Am. Chem. Soc.* **2014,** *136* (12), 4651-4658.
23. Liu, W.; Luo, X.; Bao, Y.; Liu, Y. P.; Ning, G.-H.; Abdelwahab, I.; Li, L.; Nai, C. T.; Hu, Z. G.; Zhao, D., A two-dimensional conjugated aromatic polymer via C–C coupling reaction. *Nat. Chem.* **2017,** *9* (6), 563-570.
24. Simonov, K. A.; Vinogradov, N. A.; Vinogradov, A. S.; Generalov, A. V.; Zagrebina, E. M.; Martensson, N.; Cafolla, A. A.; Carpy, T.; Cunniffe, J. P.; Preobrajenski, A. B., Effect of substrate chemistry on the bottom-up fabrication of graphene nanoribbons: combined core-level spectroscopy and STM study. *J. Phys. Chem. C* **2014,** *118* (23), 12532-12540.
25. Zhang, X.; Gärisch, F.; Chen, Z.; Hu, Y.; Wang, Z.; Wang, Y.; Xie, L.; Chen, J.; Li, J.; Barth, J. V., Self-assembly and photoinduced fabrication of conductive nanographene wires on boron nitride. *Nat. Commun.* **2022,** *13* (1), 442.
26. Wagner, A. J.; Carlo, S.; Vecitis, C.; Fairbrother, D. H., Effect of X-ray irradiation on the chemical and physical properties of a semifluorinated self-assembled monolayer. *Langmuir* **2002,** *18* (5), 1542-1549.
27. Fokin, A. A.; Tkachenko, B. A.; Gunchenko, P. A.; Gusev, D. V.; Schreiner, P. R., Functionalized nanodiamonds part I. An experimental assessment of diamantane and computational predictions for higher diamondoids. *Chem. Eur. J* **2005,** *11* (23), 7091-7101.
28. Schreiner, P. R.; Chernish, L. V.; Gunchenko, P. A.; Tikhonchuk, E. Y.; Hausmann, H.; Serafin, M.; Schlecht, S.; Dahl, J. E.; Carlson, R. M.; Fokin, A. A., Overcoming lability of extremely long alkane carbon–carbon bonds through dispersion forces. *Nature* **2011,** *477* (7364), 308-311.
29. Fokin, A. A.; Chernish, L. V.; Gunchenko, P. A.; Tikhonchuk, E. Y.; Hausmann, H.; Serafin, M.; Dahl, J. E.; Carlson, R. M.; Schreiner, P. R., Stable alkanes containing very long carbon–carbon bonds. *J. Am. Chem. Soc.* **2012,** *134* (33), 13641-13650.
30. Bieri, M.; Nguyen, M.-T.; Gröning, O.; Cai, J.; Treier, M.; Aït-Mansour, K.; Ruffieux, P.; Pignedoli, C. A.; Passerone, D.; Kastler, M.; Müllen, K.; Fasel, R., Two-Dimensional Polymer Formation on Surfaces: Insight into the Roles of Precursor Mobility and Reactivity. *J. Am. Chem. Soc.* **2010,** *132* (46), 16669-16676.
31. Lackinger, M.; Heckl, W. M., A STM perspective on covalent intermolecular coupling reactions on surfaces. *J. Phys. D: Appl. Phys.* **2011,** *44* (46), 464011.
32. Frezza, F.; Sánchez-Grande, A.; Canola, S.; Lamancová, A.; Mutombo, P.; Chen, Q.; Wäckerlin, C.; Ernst, K.-H.; Muntwiler, M.; Zema, N.; Di Giovannantonio, M.; Nachtigallová, D.; Jelínek, P., Controlling On-Surface Photoactivity: The Impact of π-Conjugation in Anhydride-Functionalized Molecules on a Semiconductor Surface. *Angew. Chem. Int. Ed.* **2024,** *63* (30), e202405983.
33. Palma, C.-A.; Samorì, P., Blueprinting macromolecular electronics. *Nat. Chem.* **2011,** *3* (6), 431-436.
34. Zhang, C.; Sun, Q.; Chen, H.; Tan, Q.; Xu, W., Formation of polyphenyl chains through hierarchical reactions: Ullmann coupling followed by cross-dehydrogenative coupling. *Chem. Commun.* **2015,** *51* (3), 495-498.
35. Ebeling, D.; Šekutor, M.; Stiefermann, M.; Tschakert, J.; Dahl, J. E.; Carlson, R. M.; Schirmeisen, A.; Schreiner, P. R., Assigning the absolute configuration of single aliphatic molecules by visual inspection. *Nat. Commun.* **2018,** *9* (1), 2420.
36. Ebeling, D.; Šekutor, M.; Stiefermann, M.; Tschakert, J.; Dahl, J. E.; Carlson, R. M.; Schirmeisen, A.; Schreiner, P. R., London dispersion directs on-surface self-assembly of [121] tetramantane molecules. *ACS nano*

**2017,** *11* (9), 9459-9466.

37. Basagni, A.; Ferrighi, L.; Cattelan, M.; Nicolas, L.; Handrup, K.; Vaghi, L.; Papagni, A.; Sedona, F.; Di Valentin, C.; Agnoli, S., On-surface photo-dissociation of C–Br bonds: towards room temperature Ullmann coupling. *Chem. Commun.* **2015,** *51* (63), 12593-12596.

38. Al‐Bataineh, S. A.; Britcher, L. G.; Griesser, H. J., Rapid radiation degradation in the XPS analysis of antibacterial coatings of brominated furanones. *Surf. Interface Anal.* **2006,** *38* (11), 1512-1518.

39. Nečas, D.; Klapetek, P., Gwyddion: an open-source software for SPM data analysis. *Open Phys.* **2012,** *10* (1), 181-188.

40. Strohalm, M.; Kavan, D.; Novák, P.; Volny, M.; Havlicek, V., mMass 3: a cross-platform software environment for precise analysis of mass spectrometric data. *Anal. Chem.* **2010,** *82* (11), 4648-4651.

41. Giannozzi, P.; Baroni, S.; Bonini, N.; Calandra, M.; Car, R.; Cavazzoni, C.; Ceresoli, D.; Chiarotti, G. L.; Cococcioni, M.; Dabo, I.; Dal Corso, A.; de Gironcoli, S.; Fabris, S.; Fratesi, G.; Gebauer, R.; Gerstmann, U.; Gougoussis, C.; Kokalj, A.; Lazzeri, M.; Martin-Samos, L.; Marzari, N.; Mauri, F.; Mazzarello, R.; Paolini, S.; Pasquarello, A.; Paulatto, L.; Sbraccia, C.; Scandolo, S.; Sclauzero, G.; Seitsonen, A. P.; Smogunov, A.; Umari, P.; Wentzcovitch, R. M., QUANTUM ESPRESSO: a modular and open-source software project for quantum simulations of materials. *J. Phys.: Condens.Matter* **2009,** *21* (39), 395502.

42. Giannozzi, P.; Andreussi, O.; Brumme, T.; Bunau, O.; Buongiorno Nardelli, M.; Calandra, M.; Car, R.; Cavazzoni, C.; Ceresoli, D.; Cococcioni, M.; Colonna, N.; Carnimeo, I.; Dal Corso, A.; de Gironcoli, S.; Delugas, P.; DiStasio, R. A.; Ferretti, A.; Floris, A.; Fratesi, G.; Fugallo, G.; Gebauer, R.; Gerstmann, U.; Giustino, F.; Gorni, T.; Jia, J.; Kawamura, M.; Ko, H. Y.; Kokalj, A.; Küçükbenli, E.; Lazzeri, M.; Marsili, M.; Marzari, N.; Mauri, F.; Nguyen, N. L.; Nguyen, H. V.; Otero-de-la-Roza, A.; Paulatto, L.; Poncé, S.; Rocca, D.; Sabatini, R.; Santra, B.; Schlipf, M.; Seitsonen, A. P.; Smogunov, A.; Timrov, I.; Thonhauser, T.; Umari, P.; Vast, N.; Wu, X.; Baroni, S., Advanced capabilities for materials modelling with Quantum ESPRESSO. *J. Phys.: Condens.Matter* **2017,** *29* (46), 465901.

43. Perdew, J. P.; Ruzsinszky, A.; Csonka, G. I.; Vydrov, O. A.; Scuseria, G. E.; Constantin, L. A.; Zhou, X.; Burke, K., Restoring the Density-Gradient Expansion for Exchange in Solids and Surfaces. *Phys. Rev. Lett.* **2008,** *100* (13), 136406.

44. Baroni, S.; de Gironcoli, S.; Dal Corso, A.; Giannozzi, P., Phonons and related crystal properties from density-functional perturbation theory. *Rev. Mod. Phys.* **2001,** *73* (2), 515-562.

45. van Setten, M. J.; Giantomassi, M.; Bousquet, E.; Verstraete, M. J.; Hamann, D. R.; Gonze, X.; Rignanese, G. M., The PseudoDojo: Training and grading a 85 element optimized norm-conserving pseudopotential table. *Comput. Phys. Commun.* **2018,** *226*, 39-54.

46. Hourahine, B.; Aradi, B.; Blum, V.; Bonafé, F.; Buccheri, A.; Camacho, C.; Cevallos, C.; Deshaye, M.; Dumitrică, T.; Dominguez, A., DFTB+, a software package for efficient approximate density functional theory based atomistic simulations. *J. Chem. Phys.* **2020,** *152* (12), 124101.

# In-architecture X-ray assisted C-Br dissociation for on-surface fabrication of diamondoid chains

Yan Wang[1,2], Niklas Grabicki[3], Hibiki Orio[4], Juan Li[2], Jie Gao[1,2], Xiaoxi Zhang[1], Tiago F. T. Cerqueira[5], Miguel A. L. Marques[6], Zhaotan Jiang[2], Friedrich Reinert[4], Oliver Dumele[3,7], Carlos-Andres Palma[1]

[1]*Institute of Physics, Chinese Academy of Sciences, 100190 Beijing, P. R. China*

[2]*School of Physics & Advanced Research Institute of Multidisciplinary Science, Beijing Institute of Technology, 100081 Beijing, P. R. China*

[3]*Department of Chemistry & IRIS Adlershof, Humboldt University of Berlin, 12489 Berlin, Germany*

[4]*Experimentelle Physik VII and Würzburg-Dresden Cluster of Excellence ct.qmat, Universität Würzburg, 97074 Würzburg, Germany*

[5]*CFisUC, Department of Physics, University of Coimbra, Rua Larga, 3004-516 Coimbra, Portugal*

[6]*Research Center Future Energy Materials and Systems of the University Alliance Ruhr and Interdisciplinary Centre for Advanced Materials Simulation, Ruhr University Bochum, Universitätsstraße 150, D-44801 Bochum, Germany*

[7]*Institute of Organic Chemistry, University of Freiburg, Albertstrasse 21, 79104 Freiburg, Germany*

# Additional information and experiment results

**Molecule synthesis**

Reagents (Acros, AlfaAesar, Sigma-Aldrich, and TCI) were purchased as reagent grade and used without further purification, unless otherwise specified. Solvents for synthesis were dried using a Pure Solv Micro Solvent Purification System from Innovative Technology and stored over molecular sieves 3-4 Å. All non-aqueous reactions were performed in oven-dried glassware and under a $N_2$ or Ar atmosphere. Automated Medium Pressure Column Chromatography (MPLC) was performed on a Teledyne ISCO CombiFlash $R_f$ 300 system with 200 mL min$^{-1}$ max flow, 200 psi, equipped with integrated ELSD and 200-800 nm UV/Vis variable wavelength detector. Thin layer chromatography (TLC) was conducted on aluminum sheets coated with $SiO_2$-60 $F_{254}$ obtained from Merck; visualization with a UV lamp (254 or 366 nm). Evaporation *in vacuo* was performed at 40-60 °C and 700–10 mbar. All products were dried under high vacuum (ca. $10^{-2}$ mbar) before analytical characterization. Reported yields refer to spectroscopically and chromatographically pure compounds that were dried under high vacuum (ca. $10^{-2}$ mbar) before analytical characterization, unless otherwise specified. Nuclear magnetic resonance (NMR) spectra were recorded using a Bruker Avance II 300 (300 MHz for $^1$H and 75 MHz for $^{13}$C) and a Bruker Avance II 500 (500 MHz for $^1$H and 126 MHz for $^{13}$C) at 298 K and are reported as follows: chemical shift ($\delta$) in ppm (multiplicity, coupling constant $J$ in Hz, number of protons; assignment). The residual deuterated solvent was used as the internal reference ($CDCl_3$: $\delta_H$ = 7.26 ppm, $CD_2Cl_2$: $\delta_H$ = 5.32 ppm, $CD_3OD$: $\delta_H$ = 3.31 ppm, $(CD_3)_2SO$: $\delta_H$ = 2.50 ppm; ($CDCl_3$: $\delta_C$ = 77.16 ppm, $CD_2Cl_2$: $\delta_C$ = 54.00 ppm, $CD_3OD$: $\delta_C$ = 49.00 ppm, $(CD_3)_2SO$: $\delta_C$ = 39.52 ppm; The resonance multiplicity is described as s (singlet), d (doublet), t (triplet), q (quartet), quint (quintet), sept. (septet), m (multiplet), and br. (broad).

## Synthetic Procedures

### 4-Bromodiamantane[1,2]

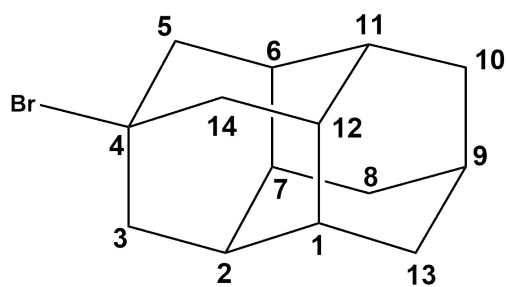

According to a modified procedure from the literature,[1,2] diamantane (1.00 g, 5.31 mmol) was dispersed together with *tert*-butyl-bromide (1.00 g 7.3 mmol) in 5 mL cyclohexane. The reaction mixture was cooled to 5 °C with an ice-water bath. AlBr$_3$ (0.05 g, 0.2 mmol) was added upon which the dispersion turned green. The reaction was kept at this temperature for 12 hours and then poured onto ice. The organic phase was separated and the aqueous phase was extracted with hexane (3 × 50 mL). The organic phases were combined, dried and the solvent was evaporated under vacuum. The crude product was purified using a preparative MPLC (SiO$_2$, neat cyclohexane). The obtained white solid was recrystallized three times in acetone to yield colorless crystals of 4-bromodiamantane (103 mg, 7%). The obtained data is in accordance with the literature.[1,2]

### 4,9-Dibromodiamantane[1,2]

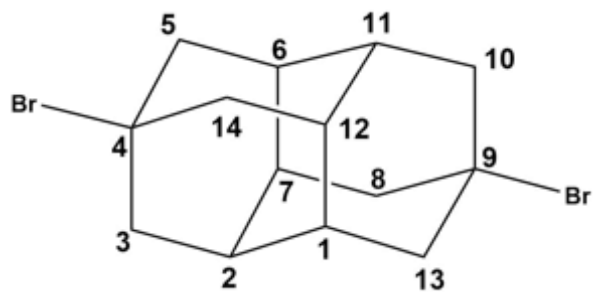

According to a modified procedure from the literature[1,2], AlBr$_3$ (453 mg, 1.7 mmol) was suspended in a cyclohexane/pentane 5:1 (30 mL), diamantane (4.00 g, 21.2 mmol) was added and the reaction mixture was cooled down to −10 °C. The suspension was stirred for 5 minutes. Then bromine was added dropwise (17.7 g, 110 mmol, 3 drops per minute), keeping the temperature at −7 °C. The reaction mixture was allowed to warm up to 10 °C within 2 hours and then quenched with aq. sat.

NaS$_2$O$_3$ (ca. 25 mL) until the orange color disappeared. The aqueous phase was extracted with hexane (1×40 mL) and CH$_2$Cl$_2$ (2×40 mL). The combined organic phases were washed with water (1×20 mL), and dried over MgSO$_4$. The solvent was removed under reduced pressure. The obtained dark solid was heated in acetone (350 mL) and filtered while hot and evaporated to a total volume of 55 mL. The solution was cooled down to 8 °C in the fridge and a colorless precipitate was collected by filtration (1.4 g). This solid was recrystallized two more times from acetone to yield **4,9-dibromodiamantane** (637 mg, 7%) as colorless crystals. The obtained data is in accordance with the literature.[1,2] Elem. anal. calcd. (%) for C$_{14}$H$_{18}$Br$_2$: C 48.58; H 5.24, found: C 48.87, H 5.25.

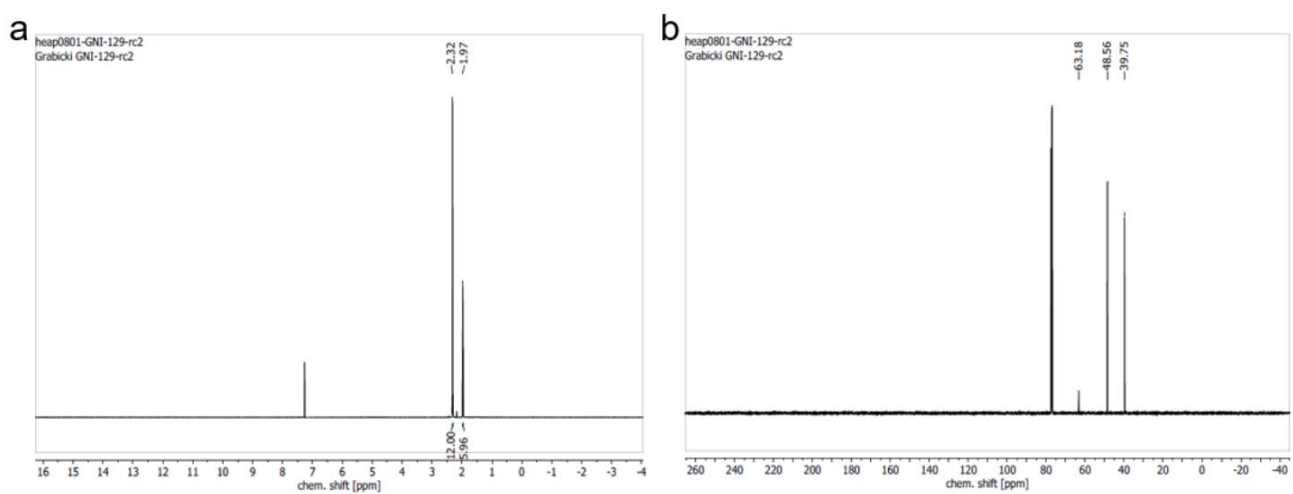

**Figure S1. NMR spectra of 4,9-dibromodiamantane in CDCl3. a**. $^1$HNMR spectrum of 4,9-dibromodiamantane, signals are responding to aliphatic singlets. **b**. $^{13}$C NMR spectrum of 4,9-dibromodiamantane molecule.

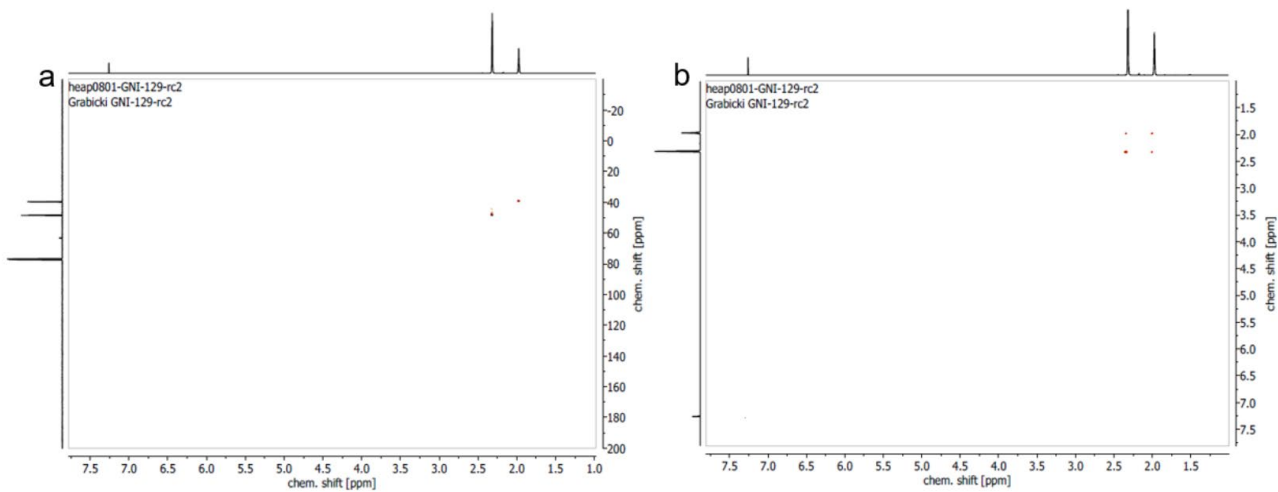

**Figure S2. 2D NMR spectrum of 4,9-dibromodiamantane. a**. $^1$H,$^{13}$C HSCQ NMR spectrum of 4,9-dibromodiamantane. **b**. $^1$H,$^1$H COSY NMR spectrum of 4,9-dibromodiamantane.

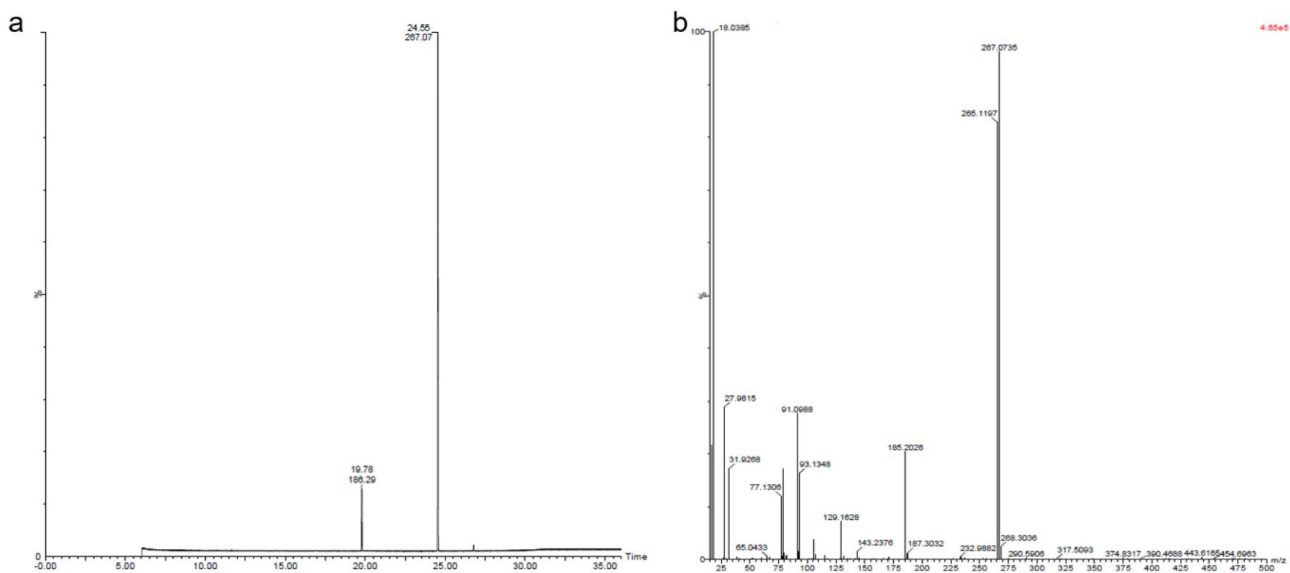

**Figure S3. GC chromatogram and mass spectrum of 4,9-dibromodiamantane. a**. GC chromatogram of a 4,9-dibromodiamantane. **b**. Mass spectrum of a 4,9-dibromodiamantane sample (calcd. $m/z$ for $M^+$: 343.98, calcd. $m/z$ for [$M$–Br = $C_{14}H_{18}{}^{79}Br^{•+}$]: 265.06, and calcd. $m/z$ for [$M$–Br = $C_{14}H_{18}{}^{81}Br^{•+}$]: 267.06, calcd. $m/z$ for [$M$–2Br+H = $C_{14}H_{18}{}^{(•••)+}$]: 186.14).).

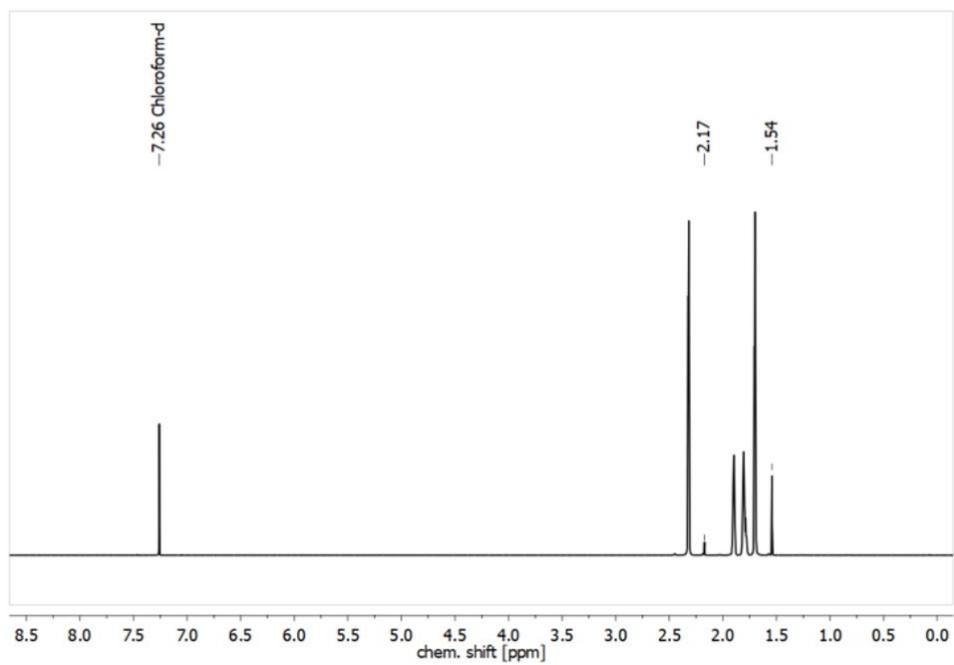

**Figure S4.** $^1$H NMR spectrum of 4-bromodiamantane in CDCl$_3$.

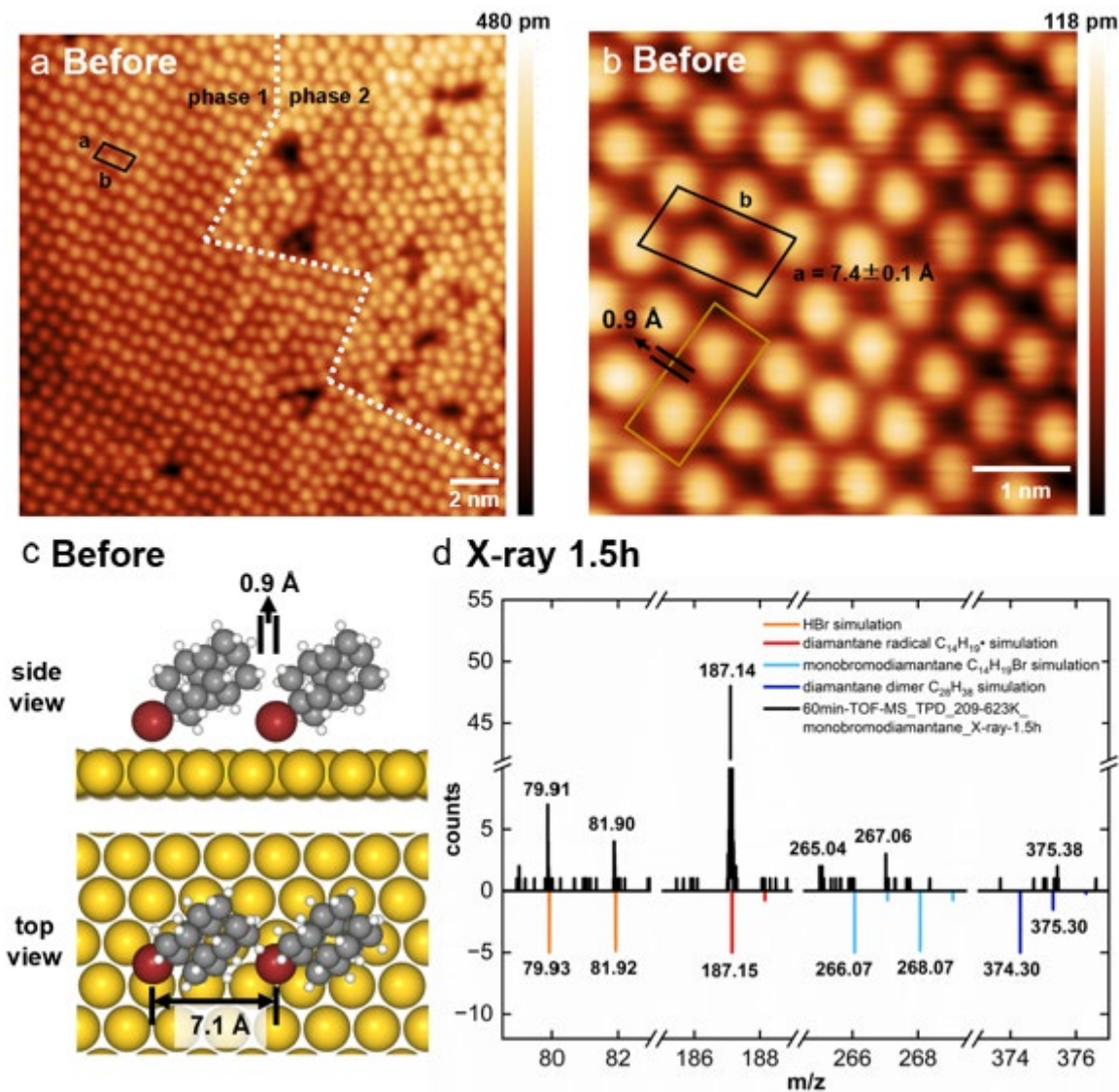

**Figure S5. Monobromodiamantane molecules on Au(111) before and after X-ray irradiation. a**. STM image of the self-assembly of monobromodiamantane molecules on Au(111), the unit cell is **a** = 7.4 ± 0.1 Å, **b** = 13.4 ± 0.1Å (0.8 V, 20 pA). **b.** Magnified STM image of the self-assembly on Au(111) (0.8 V, 20 pA). **c**. Proposed spatial configurations of monobromodiamantane molecules on Au(111), corresponding to the molecules marked by yellow rectangle in panel b. **d**. TOF-MS spectra of monobromodiamantane molecules after X-ray irradiation for 1.5 hours at sample temperature of 218 K, and the simulation of HBr, diamantane radical, monobromodiamantane, and diamantane dimer.

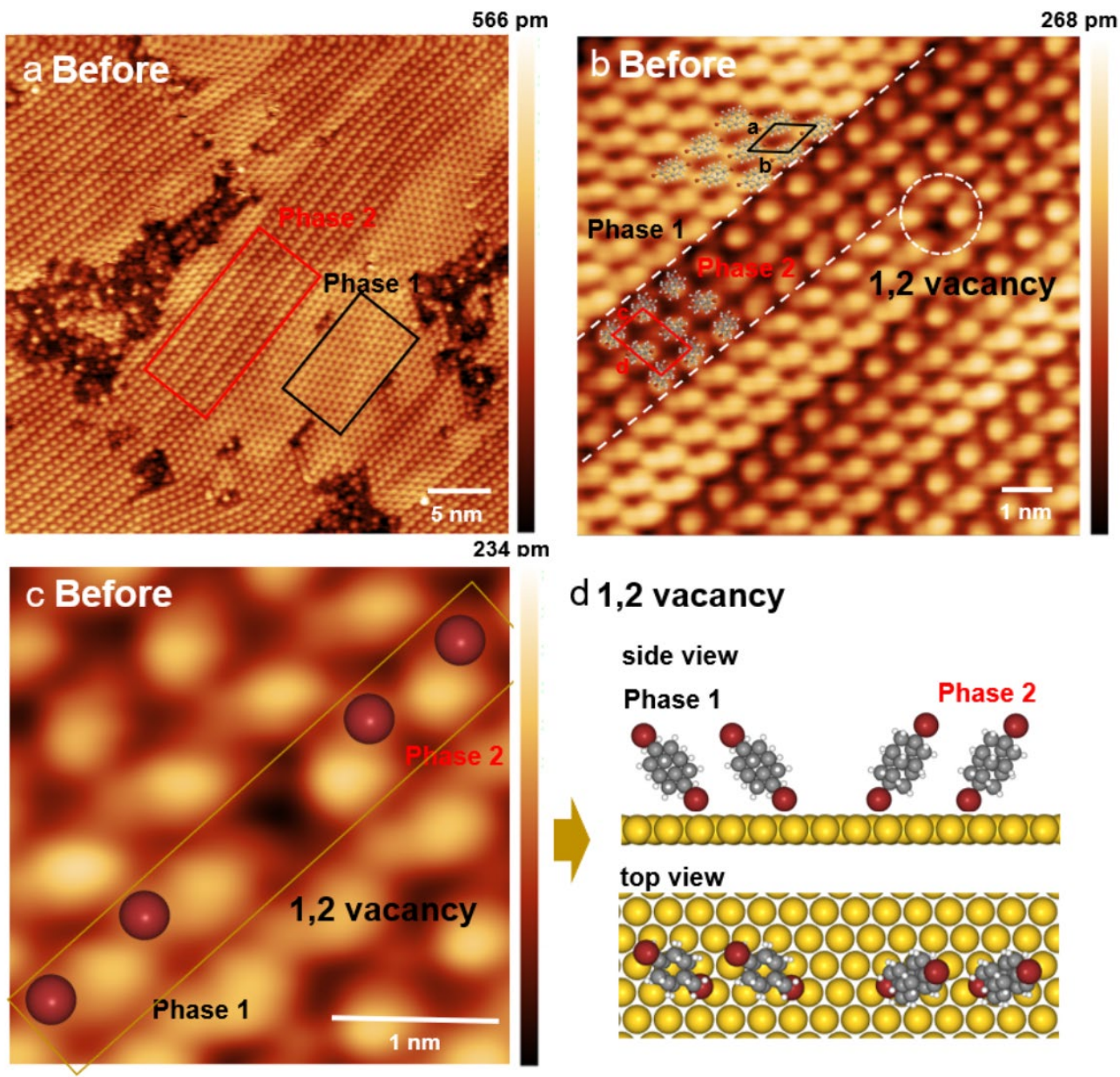

**Figure S6. STM data of dibromodiamantane on Au(111). a**. Overview STM image of dibromodiamantane on Au(111) shown two different configurations (800 mV, 35 pA). **b**. Magnified STM image of dibromodiamantane on Au(111). **c**. Zoomed-in STM image of 1.2 vacancy. **d**. Proposed configurations of dibromodiamantane on Au(111) surface.

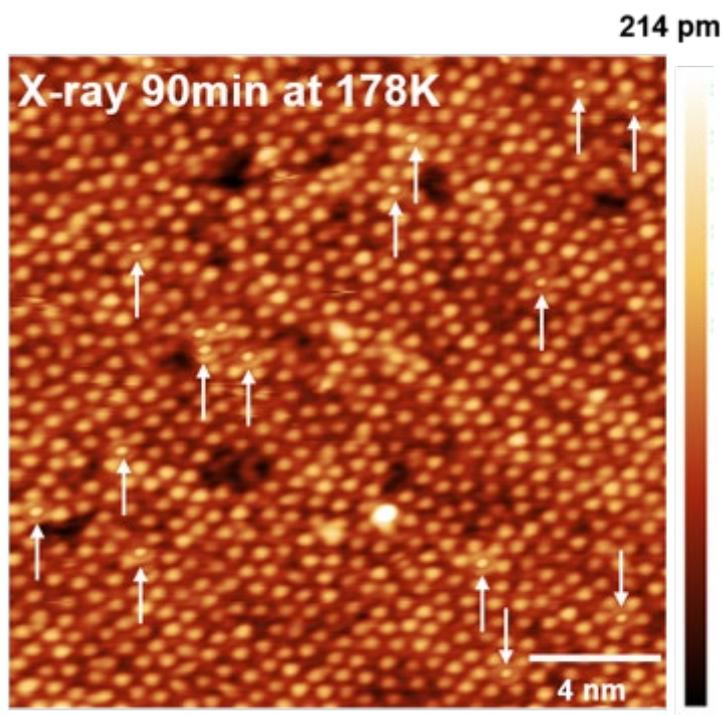

**Figure S7. STM data of bromodiamantane on Au(111) after X-ray irradiation at 178 K.** The sharp dots are uniquely observed after X-ray irradiation and could arise from contaminants or interaction of debrominated monomers with the gold surface.

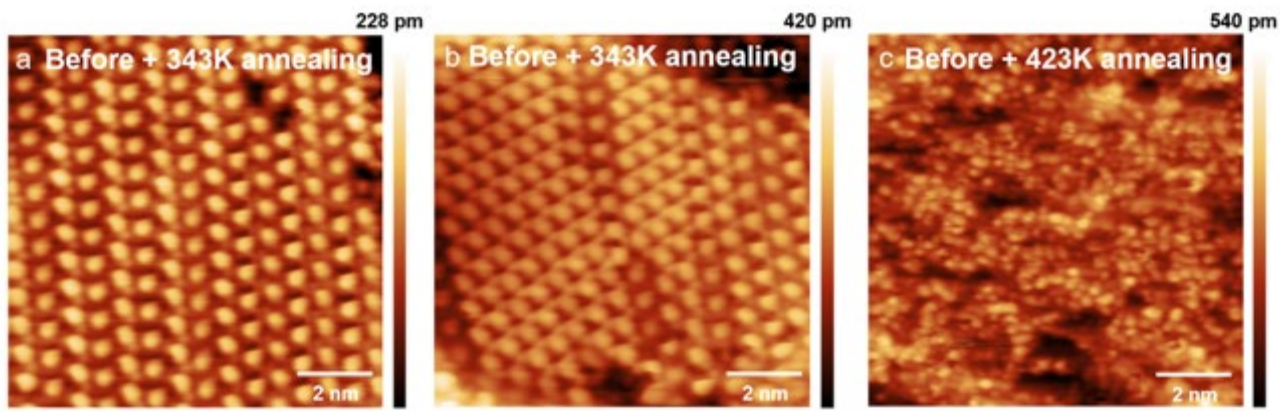

**Figure S8. STM data of dibromodiamantane on Au(111) without X-ray after annealing to 343K, 423K. a-b**. STM data of dibromodiamantane on Au(111) after annealing to 343 K. **c**. STM image of dibromodiamantane on Au(111) after annealing to 423 K.

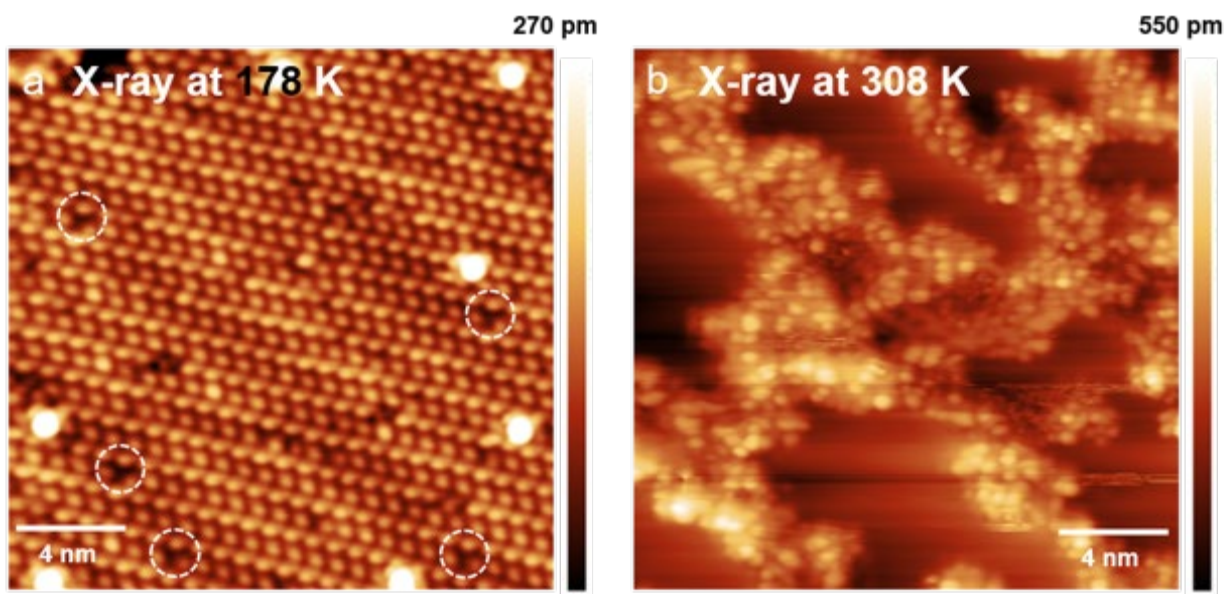

**Figure S9. STM data of dibromodiamantane on Au(111) after X-ray irradiation up to 308 K. a**. STM data of dibromodiamantane on Au(111) after X-ray irradiation at 178 K. **b**. STM data of dibromodiamantane on Au(111) after X-ray irradiation at 308 K.

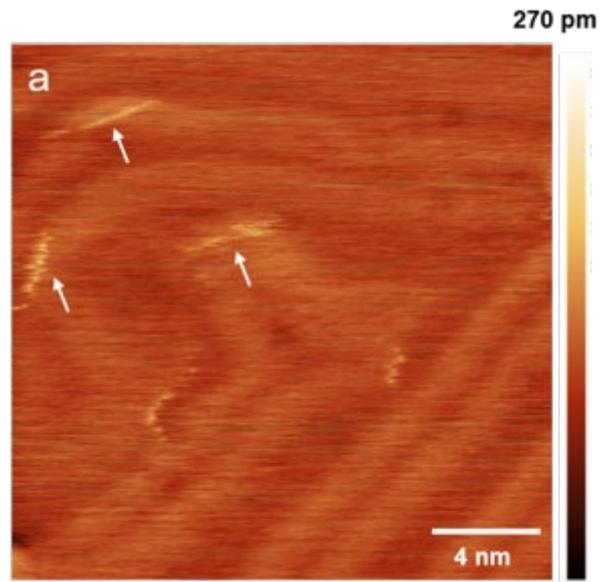

**Figure S10. RT-STM image of sample after X-Ray irradiation.** Scanning parameters: $V_b$ = 1 V, $I_t$ = 25 pA.

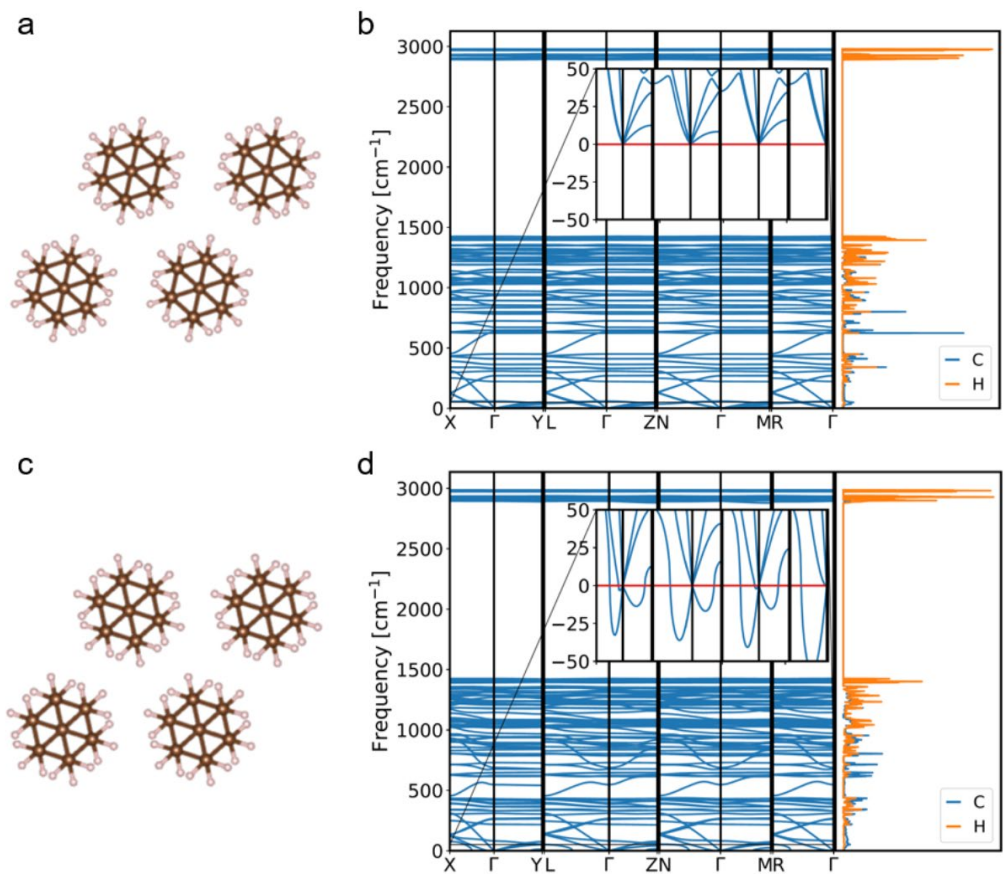

**Figure S11. The simulation of polydiamantane materials. a-b.** The structure (a) and the phonon spectra (b) of polydiamantane. **c-d.** The structure (c) and the phonon spectra (d) of hole-doped polydiamantane.

**References:**

[1] T. M. Gund, M. Nomura, P. v. R. Schleyer, *J. Org. Chem.* **1974**, *39*, 2995–3003
[2] T. M. Gund, M. Nomura, V. Z. Williams Jr., P.v.R.Schleyer, C.Hoogzand, *Tetrahedron Lett.,* **1970**, *11*, 4875-4878